\documentstyle[prb,aps,multicol,epsfig]{revtex}

\begin{document}

\title{Hopping Transport in the Presence of Site Energy Disorder:\\
       Temperature and Concentration Scaling of Conductivity Spectra}

\author{Markus~Porto$^{1,}$\footnote{New affiliation:
                                     School of Chemistry,
                                     Tel Aviv University,
                                     69978 Tel Aviv, Israel},
        Philipp Maass$^2$, Martin Meyer$^3$,
        Armin~Bunde$^1$, and~Wolfgang~Dieterich$^2$\vspace*{0.1cm}}

\address{$^1$Institut f\"ur Theoretische Physik~III,
         Justus-Liebig-Universit\"at~Giessen,
         Heinrich-Buff-Ring~16, 35392~Giessen, Germany\vspace*{0.05cm}}
\address{$^2$Fakult\"at f\"ur Physik, Universit\"at Konstanz,
         Universit\"atsstr.~10, 78457~Konstanz, Germany\vspace*{0.05cm}}
\address{$^3$Departement of Physics, Boston University, USA\vspace*{0.05cm}}

\date{August 27, 1999; revised November 5, 1999}

\maketitle

\begin{abstract}
  Recent measurements on ion conducting glasses have revealed that conductivity
  spectra for various temperatures $T$ and ionic concentrations $n$ can be
  superimposed onto a common master curve by an appropriate rescaling of the
  conductivity and frequency. In order to understand the origin of the observed
  scaling behavior, we investigate by Monte Carlo simulations the diffusion of
  particles in a lattice with site energy disorder for a wide range of both $T$
  and $n$ values. While the model can account for the changes in ionic
  activation energies upon changing $n$, it in general yields conductivity
  spectra that exhibit no scaling behavior. However, for typical $n$ and
  sufficiently low $T$ values, a fairly good data collapse is obtained
  analogous to that found in experiment.
\end{abstract}

\pacs{PACS numbers: 66.30.Dn, 66.30.Hs}

\begin{multicols}{2}

\section{Introduction}\label{intro-sec}

Electrical conduction processes in disordered solids show a remarkable
similiarity with respect to their low-frequency dielectric loss behavior.
\cite{Jonscher:1977} For frequencies $\omega$ smaller than some crossover
frequency $\tau^{-1}$, the real part $\sigma'(\omega)$ of the complex
conductivity $\hat\sigma(\omega) = \sigma'(\omega)+i\sigma''(\omega)$ is
constant ($\sigma'(\omega) = \sigma(0) = \sigma_{\rm dc}$ for $\omega\tau \ll
1$), while for higher frequencies $\omega\tau > 1$, $\sigma'(\omega)$ increases
monotonously with $\omega$, until at typical phonon frequencies above
$100\;{\rm GHz}$ vibrational contributions become dominant. The dispersive part
can be characterized by {\it approximate\/} power laws, $\sigma'(\omega) \sim
\omega^s$, with exponents $s \simeq 0.6-0.8$, where $s$ tends to increase
weakly with increasing frequency (at fixed temperature $T$) and decreasing
temperature. In fact, at low temperatures or high frequencies the exponents $s$
becomes close to one. Alternatively, this general behavior may be discussed in
terms of the complex dielectric function $\hat\epsilon(\omega) =
\epsilon'(\omega) + i\epsilon''(\omega) = \epsilon_{\infty} +
i\hat\sigma(\omega)/\epsilon_0\epsilon_{\infty}$, where $\epsilon_0$ is the
vacuum permittivity and $\epsilon_{\infty}$ the high frequency dielectric
constant.

For ionically conducting glasses in particular, it was pointed out by Isard
\cite{Isard:1962} already in 1962 that $\tau^{-1}$ and $\sigma(0)$ are
proportional with an almost universal constant of proportionality. More
precisely, it was found that the peak frequency $\omega_p$ in the ``dielectric
polarization part'' $[\sigma'(\omega)-\sigma(0)]/\omega$ of the loss spectrum
is proportional to $\sigma(0)$. However, since any reasonable definition of
$\tau^{-1}$ yields values comparable to $\omega_p$, we do not distinguish
between $\omega_p$ and $\tau^{-1}$ here. A closer inspection of the relation
between $\tau^{-1}$ and $\sigma(0)$ was done later independently by Barton,
Nakajima, and Namikawa \cite{Barton:1966+Nakajima:1971+Namikawa:1975} on a
variety of ionic glasses with the result
\begin{equation}\label{eq:bnn}
\sigma(0) \tau = p \epsilon_0 \Delta\epsilon'{\;},
\end{equation}
where $p$ is a material dependent constant of order one and $\Delta\epsilon' =
\epsilon'(0)-\epsilon_{\infty} = -\epsilon_0^{-1} \lim_{\omega \to 0}
\sigma''(\omega)/\omega$ is the dielectric loss strength.
Equation~(\ref{eq:bnn}) is commonly referred to as the BNN relation.

The question if and to what extent these common features of conductivity
spectra in disordered solids can be cast into a universal scaling form has been
debated for a long time. In fact, if $\tau$ is the only relevant time scale in
the system and $\Delta\epsilon' \propto n/T$ with $n$ being the charge carrier
concentration (this is the simplest behavior one can expect based on the
fluctuation-dissipation theorem), the BNN relation suggests a scaling form
\begin{equation}\label{eq:bnn-scaling}
\frac{\hat\sigma(\omega)}{\sigma(0)} = F(\omega\tau){\;}, \qquad
\tau = \frac{\alpha n}{\sigma(0) T} {\;}\,,
\end{equation}
where $\alpha$ is a constant of order unity. This type of scaling behavior will
henceforth be referred to as the BNN scaling.

Various theories of charge carrier transport in disordered systems have been
developed over the past 30 years, and already within the classical pair
approximation \cite{Pollak+Geballe:1961} and the continuos time random walk
(CTRW) model \cite{Scher/Lax:1973+Odagaki/Lax:1981} the possibility of scaling
was considered. The problem has been investigated in more detail then by a
number of authors. Summerfield suggested, \cite{Summerfield:1985} based on the
extended pair approximation (EPA), \cite{Summerfield/Butcher:1982} that the
real part $\sigma'(\omega)$ in doped crystalline and amorphous semiconductors
should follow eq.~(\ref{eq:bnn-scaling}) (but without considering the
dependence of $\tau$ on the charge carrier concentration $n$). Tests of this
scaling ansatz for data on evaporated amorphous germanium (for $51{\;}{\rm K} <
T < 102{\;}{\rm K}$) and doped crystalline silicon (for $1{\;}{\rm K} < T <
10{\;}{\rm K}$) in a frequency range $10^{-3}-10^5{\;}{\rm kHz}$, showed a good
data collapse for various temperatures and indeed almost the same scaling
function $F(\omega\tau)$ for doped crystalline silicon and amorphous germanium.
\cite{Note1} Hunt \cite{Hunt:1991+Hunt:1992+Hunt:1994} suggested a scaling
behavior for the real part $\sigma'(\omega)$ using an ``augmented pair
approximation'' at intermediate frequencies and a ``cluster theory'' at low
frequencies, and discussed the degree of universality of the BNN relation.

Dyre \cite{Dyre:1993+Dyre:1993} studied a special electric circuit model
resembling a discretized form of Maxwell's equations for a medium with
spatially fluctuating dielectric function. By treating this model in the
effective medium approximation (EMA), he showed that at low temperatures the
complex conductivity $\hat\sigma(\omega)$ can be written as
\begin{equation}\label{eq:bryksin}
\frac{\hat\sigma(\omega)}{\sigma(0)}
\log\left(\frac{\hat\sigma(\omega)}{\sigma(0)}\right) = i\omega\tau{\;},
\end{equation}
where $\tau \propto 1/T \sigma(0)$. This low temperature limit of the EMA was
first derived by Bryksin \cite{Bryksin:1980} and verified to hold true for
various distributions of the impedances in the electric circuit model,
\cite{Dyre:1993+Dyre:1993} thus providing a possible explanation for the
experimentally observed ``quasi-universality''. Bryksin's equation
(\ref{eq:bryksin}) also results from an EMA treatment of a microscopic hopping
model \cite{Fishchuk:1986,Dyre:1988} with uniformly distributed random free
energy barriers at low $T$ (see also Ref. \cite{Schirmacher:1988}). This
treatment yields a crossover time $\tau \propto n/T \sigma(0)$, in accordance
with the BNN scaling (\ref{eq:bnn-scaling}). Note that Bryksin's equation
predicts a unique scaling function $F(z)$ being determined by $F(z) \log F(z) =
iz$.

The Bryksin scaling function was shown to describe the normalized real part
$\sigma'(\omega)/\sigma(0)$ of various materials fairly well, \cite{Note2} both
for electronic conductors (as e.g.\ n-doped crystalline silicon or amorphous
germanium) and ionic conductors (as e.g.\ sodium silicate glass or the glass
former $0.4{\rm Ca}{\rm N}{\rm O}_3 \cdot 0.6{\rm K}{\rm N}{\rm O}_3$). It must
be noted, however, that the EMA fails to predict the correct dc-conductivity
$\sigma(0)$. The low temperature activation energy $E_{\rm a} \propto
-\log(T\sigma(0))$ resulting from the EMA is not in accordance with the value
obtained from a critical percolation path analysis,
\cite{Ambegoakar/etal:1971+Shklovskii/Efros:1971} which is known to become
exact at low temperatures \cite{Tyc/Halperin:1989} (see also the results in
Sec.~\ref{section:results} below). Only with a rescaled dc-conductivity may the
EMA be regarded as a valuable concept to describe conductivity spectra in
strongly disordered systems.

For a long time the BNN scaling (\ref{eq:bnn-scaling}) has never been tested
with respect to changes in charge carrier concentration $n$, until Roling {\it
et al.\/} recently studied $\hat\sigma(\omega)$ for various ionically
conducting glasses \cite{Roling/etal:1997,Roling:1998} (see also the recent
study of Sidebottom \cite{Sidebottom:1999}). In these materials, in contrast to
most other systems, the concentration of charge carriers (mobile ions) can be
varied to a large extent. Roling {\it et al.} found that the {\it real parts\/}
of conductivity spectra in sodium borate glasses with compositions $x {\rm
Na}_2{\rm O} \cdot (1-x) {\rm B}_2{\rm O}_3$ (and also other types of ionic
glasses \cite{Roling:1998}) can be superimposed fairly well onto a single curve
\cite{Note3} not only for various temperatures, but also for various modifier
contents $x \sim n$. However, it was found also \cite{Roling:1998} that the
dielectric strength scales as $\Delta\epsilon' \propto x^{1/3} \sim n^{1/3}$,
in contrast to the assumption $\Delta\epsilon' \propto n/T$ made when deriving
the BNN scaling (\ref{eq:bnn-scaling}). This implies that the imaginary part
cannot obey the BNN scaling because of causality requirements (Kramers-Kronig
relations). Moreover, in Ref. \cite{Sidebottom:1999} alkali germanate glasses
have been studied over $n$ values varying by a factor of ten, and it was found
that the BNN scaling has to be modified in order to obtain a data collapse for
this large concentration range. We therefore have to consider the observed data
collapse in $\sigma(\omega)$ as an effective one.

The effective scaling behavior is nevertheless quite remarkable, since the
activation energy $E_{\rm a}$ of the dc-conductivity $\sigma(0)$ (and hence the
activation energy for $\tau^{-1} \propto \sigma(0)$) usually decreases
logarithmically \cite{Maass/Bunde/Ingram:1992,Pradel/Ribes:1994} with ionic
concentration $n$, $E_{\rm a} = A-B \log n$, for typical $n\approx
10^{21}-10^{22} \; {\rm cm}^{-3}$. Accordingly, one finds a quite complicated
explicit dependence of the crossover time $\tau$ on $n$, $\tau \sim
n^{1-B/k_{\rm B} T}$.

It is important to note that these experimental findings for ionically
conducting glasses cannot be explained by the random free energy barrier model,
since in this model the conductivity is proportional to $n$ but does not
steeply increase with $n$ as $\sigma(0) \sim n^{B/k_{\rm B} T}$. The perhaps
simplest approach that allows one to account for a concentration dependence of
$E_{\rm a}$ is a model of particles hopping in a lattice with {\it site energy
disorder\/} (see Refs. \cite{Hunt:1994,Kehr:1998,Maass:1999} and below), where
only one particle can occupy a given lattice site (Fermi statistics). In order
to see if such a model also yields the BNN scaling (\ref{eq:bnn-scaling}), we
perform Monte Carlo simulations for Gaussian distributed site energies in a
range of concentrations $n$ and temperatures $T$ both extending over almost two
orders of magnitude. We find that the model can account for the logarithmic
dependence of $E_{\rm a}$ on $n$, but in general exhibits no effective BNN
scaling behavior corresponding to (\ref{eq:bnn-scaling}). However, for typical
ionic concentrations and sufficiently low temperatures an effective BNN scaling
behavior is found in agreement with experiment.

\section{Hopping Motion and\\
\hspace*{0.8cm}Transport Quantities}\label{section:model}

The jump motion of particles in a lattice with site energy disorder is a
standard model for describing hopping transport in disordered systems.
\cite{Boettger/Bryksin:1985} We choose here a simple cubic lattice with spacing
$a$ and assign to each lattice site $i$ independently a random energy
$\varepsilon_i$ drawn from a Gaussian distribution $\psi(\varepsilon)$ with
zero mean and variance $\sigma_{\varepsilon}^2$. The particles jump among
nearest neighbor sites and the jump rate from site $i$ to a vacant neighboring
site $j$ is $W(\varepsilon_i,\varepsilon_j) = \nu
\min(1,\exp[-(\varepsilon_j-\varepsilon_i)/k_{\rm B}T])$ with $\nu$ being an
attempt frequency (Metropolis transition rates). If the neighboring site $j$ is
occupied, the jump rate is zero. The size of the lattice is $L^3$ and periodic
boundary conditions are used. At equilibrium, the probability for a site with
energy $\varepsilon$ to be occupied is given by the Fermi distribution
$f(\varepsilon) = [1+\exp[(\varepsilon-\mu)/k_{\rm B}T]^{-1}$, where the
concentration $c = na^3$ per lattice site determines the chemical potential
through $c = (a/L)^3\sum_{i}f(\varepsilon_i)\cong\int d\varepsilon
\psi(\varepsilon)f(\varepsilon)$. For $T\to0$, $f(\varepsilon)$ approaches the
step function $f(\varepsilon) = \theta(\varepsilon_{\rm f}(c)-\varepsilon)$
($\theta(x) = 1$ for $x\ge0$ and zero else) with the Fermi energy
$\varepsilon_{\rm f}(c)$ given by
\begin{equation}
{\rm erf}(\varepsilon_{\rm f}(c)/2^{1/2}\sigma_{\varepsilon}) = 2c-1\,,
\end{equation}
where ${\rm erf}(\cdot)$ denotes the error function.

In order to analyse the mobility of the particles for various concentrations
$c$ and temperatures $T$ we determine the mean square displacement $\langle
r^2(t)\rangle$ of a tracer particle by means of Monte Carlo simulations,
\cite{Binder/Heermann:1992} and calculate the time-dependent diffusion
coefficient $D(t) = \langle r^2(t)\rangle/6t$. $D(t)$ approaches the short-time
diffusion constant $D_{\rm st}$ for $t\to0$, and the long-time diffusion
constant $D_{\infty}$ for $t\to\infty$. The frequency-dependent conductivity
$\hat\sigma(\omega)$ is, to a good approximation, \cite{Note4} given by
\begin{equation}\label{sigma-eq}
\hat{\sigma}(\omega) = -\frac{n e^2}{k_{\rm B} T}
\frac{\omega^2}{6} \lim_{\epsilon \to 0^{+}} \int_0^{\infty} \left< r^2(t)
\right> \; \exp(i \omega t - \epsilon t) \; {\rm d}t\,,
\end{equation}
and approaches $\sigma(0) = \frac{n e^2}{k_{\rm B} T} D_{\infty}$ and
$\sigma(\infty) = \frac{n e^2}{k_{\rm B} T}D_{\rm st}$ in the low and high
frequency limits, respectively.

The short-time or high-frequency dynamics is governed by the mean jump rate of
a particle,
\begin{equation}\label{w-eq}
\langle W\rangle = \int d\varepsilon\,\frac{\psi(\varepsilon)f(\varepsilon)}{c}
\int d\varepsilon'\,\psi(\varepsilon')\left[1-f(\varepsilon')\right]
W(\varepsilon,\varepsilon')\,,
\end{equation}
which determines $D_{\rm st} = \langle W\rangle a^2/6$. For $k_{\rm
B}T\gg|\varepsilon_{\rm f}(c)|$ we can use the high temperature expansion of
the chemical potential, $\mu = k_{\rm
B}T[\log[c/(1-c)]+(c-1/2)(\sigma_{\varepsilon}/k_{\rm B}T)^2+
O(\sigma_{\varepsilon}/k_{\rm B}T)^3)$ to evaluate eq.~(\ref{w-eq}) and obtain
\begin{equation}\label{whight-eq}
\langle W\rangle\simeq
\nu(1-c)\left(1-\frac{\sigma_{\varepsilon}}{\pi^{1/2}k_{\rm B}T}\right)\,.
\end{equation}
At low temperatures $k_{\rm B}T\lesssim |\varepsilon_{\rm f}(c)|$ we can
replace $f(\varepsilon)$ by the step function in eq.~(\ref{w-eq}) and it
follows that $\langle W\rangle = (4\nu/c)\,\exp[(\sigma_{\varepsilon}/k_{\rm
B}T)^2]\,{\rm erfc}E_-\,{\rm erfc}E_+$ with $E_{\pm} =
2^{-1/2}(\sigma_{\varepsilon}/k_{\rm B}T\pm \varepsilon_{\rm
f}(c)/\sigma_{\varepsilon})$ (${\rm erfc}(\cdot) = 1-{\rm erf}(\cdot)$ denotes
the complementary error function).  For $k_{\rm B}T\ll {\rm
min}(\sigma_{\varepsilon}^2/|\varepsilon_{\rm f}(c)|,|\varepsilon_{\rm f}(c)|)$
in particular, we then find, by using ${\rm erfc}(x)\sim \exp(-x^2)/\pi^{1/2}x$
for $x\to+\infty$,
\begin{equation}\label{wlowt-eq}
\langle W\rangle\simeq \frac{\nu}{2\pi
c}\left(\frac{k_{\rm B}T}{\sigma_{\varepsilon}}\right)^2
\exp[-\varepsilon_{\rm f}(c)^2/\sigma_{\varepsilon}^2]\,.
\end{equation}
Note that $\langle W\rangle$ is not thermally activated in this low temperature
limit. This is a desirable feature of the model, since the high frequency
conductivity is indeed only weakly temperature dependent in ionically
conducting glasses (and other disordered hopping systems).

By contrast, the long-time diffusion constant or dc-conductivity is thermally
activated at low temperatues, $D_{\infty}\propto\sigma(0)\propto\exp(-E_{\rm
a}/k_{\rm B}T)$. The activation energy $E_{\rm a}$ can, at sufficiently low $c$
($[\psi(\varepsilon_{\rm f}(c))k_{\rm B}T]^{1/3}\ll1$), be calculated from the
critical path analysis \cite{Ambegoakar/etal:1971+Shklovskii/Efros:1971} as the
difference between the critical energy $\varepsilon_{\rm c}$ and the Fermi
energy $\varepsilon_{\rm f}(c)$, i.e.\ $E_{\rm a} = \varepsilon_{\rm c} -
\varepsilon_{\rm f}(c)$. Here, $\varepsilon_{\rm c}$ is the highest energy,
which at least has to be encountered by a particle when it wants to move
between two (far distant) sites close to the Fermi level. According to
percolation theory \cite{Stauffer/Aharony:1992,Bunde/Havlin:1996}
$\int_{-\infty}^{\varepsilon_{\rm c}} \psi(\varepsilon) \; {\rm d}\varepsilon
=  [1+{\rm erf}(\varepsilon_{\rm c})]/2 = p_{\rm c}$, where $p_{\rm c}\cong
0.3117$ denotes the percolation threshold in the sc-lattice. Solving
numerically for $\varepsilon_{\rm c}$ we obtain
$\varepsilon_c\cong-0.491\sigma_{\varepsilon}$ and hence
\begin{equation}\label{ea-eq}
E_{\rm a}\simeq-0.491\sigma_{\varepsilon}-\varepsilon_{\rm f}(c)\,.
\end{equation}
For $c\to0$, $\varepsilon_{\rm f}(c)\to-\infty$, such that the implicit
equation ${\rm erf}(\varepsilon_{\rm f}(c)/2^{1/2}\sigma_{\varepsilon}) = 2c-1$
for $\varepsilon_{\rm f}(c)$ can be simplified to $\exp(-[\varepsilon_{\rm
f}(c)/(2^{1/2}\sigma_{\varepsilon})]^2)\simeq - 2\pi^{1/2}c\,[\varepsilon_{\rm
f}(c)/(2^{1/2}\sigma_{\varepsilon})]$, which yields $\varepsilon_{\rm
f}(c)/\sigma_{\varepsilon}\simeq
-2^{1/2}[-\log(2\pi^{1/2}c[-\log(2\pi^{1/2}c)]^{1/2})]^{1/2}$.

\section{Results and Discussion}\label{section:results}

\begin{figure}[b]
\vspace*{-4cm}
\hspace*{-1cm}\epsfig{file=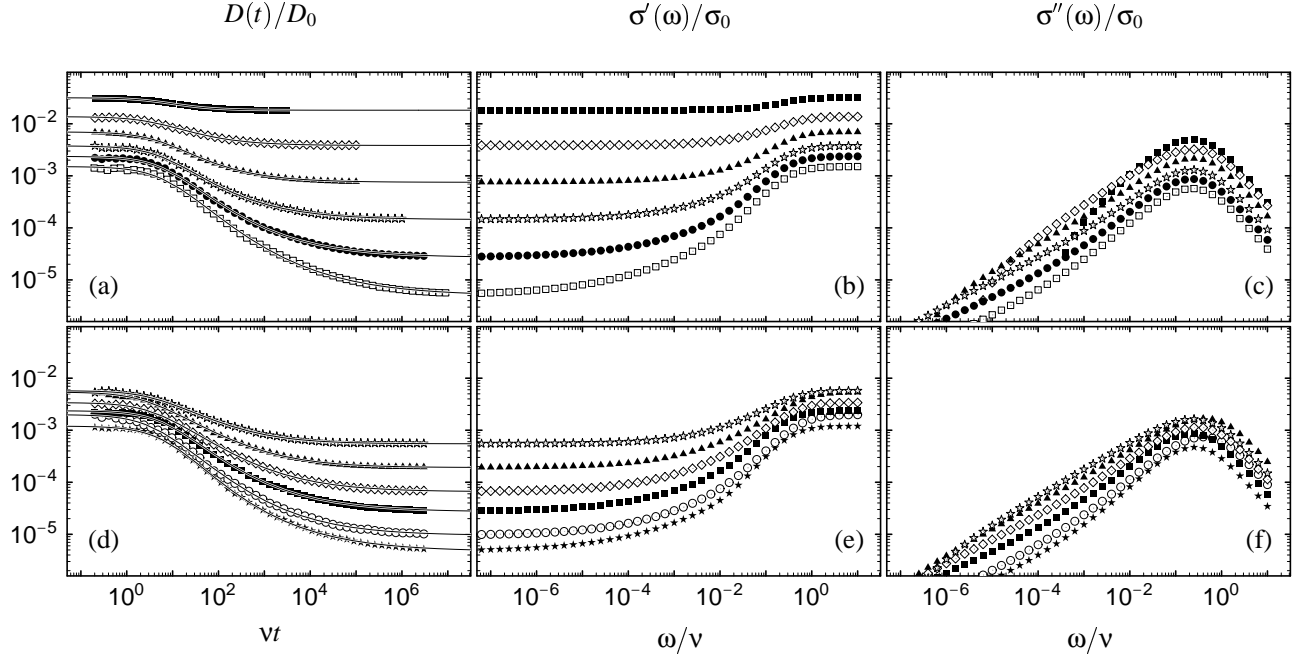,height=20cm,angle=-90}
\vspace*{-3cm}
\caption{Plot of
  (a),(d)~$D(t)/D_0$ vs $\nu t$, (b),(e)~$\sigma'(\omega)/\sigma_0$ vs
  $\omega/\nu$, and (c),(f)~$\sigma''(\omega)/\sigma_0$ vs
  $\omega/\nu$ ($\sigma_0 = ne^2D_0/k_{\rm B} T$). In (a-c)~the
  concentration is fixed, $c = 0.02$, and the temperature is varied,
  $\sigma_{\varepsilon}/k_{\rm B} T = 2$ (full squares), $3$ (open
  diamonds), $4$ (full triangle), $5$ (open stars), $6$ (full
  circles), and $7$ (open squares), while in (d-f)~the temperature is
  fixed, $\sigma_{\varepsilon}/k_{\rm B} T = 6$, and the concentration
  is varied, $c = 0.005$ (full stars), $0.01$ (open circles), $0.02$
  (full squares), $0.04$ (open diamonds), $0.08$ (full triangles), and
  $0.16$ (open stars). The solid lines in (a) and (d) represent the
  fits used for the Laplace transform.}
\label{figure:sqrr+sigma}
\end{figure}

Figure~\ref{figure:sqrr+sigma} shows the Monte Carlo results for
$D(t)$ (in units of $D_0\equiv \nu a^2$) and the real and imaginary
parts $\sigma'(\omega)$ and $\sigma''(\omega)$ of the conductivity (in
units of $\sigma_0\equiv ne^2D_0/k_{\rm B} T$) being calculated from
$D(t)$ by numerical a Laplace transformation according to
eq.~(\ref{sigma-eq}).  In the upper figures
(Figs.~\ref{figure:sqrr+sigma}a-c), the curves refer to a fixed
concentration $c = 0.02$ and various (inverse) temperatures ranging
from $\sigma_{\varepsilon}/k_{\rm B}T = 2$ to $7$, while in the lower
figures (Figs.~\ref{figure:sqrr+sigma}d-f) the temperature is held
fixed, $\sigma_{\varepsilon}/k_{\rm B}T = 6$, and the concentration is
varied between $c = 0.005$ and $c = 0.16$.  The overall behavior of
the complex conductivity $\hat\sigma(\omega)$ compares fairly well
with that found in the experiments.  There exists a crossover time
$\tau$, such that the real part $\sigma'(\omega)$ is almost constant
for $\omega\tau\ll1$, $\sigma'(\omega)\simeq\sigma(0)$. For
$\omega\tau>1$, $\sigma'(\omega)$ increases monotonously with $\omega$
until for $\omega\simeq\omega_{\rm hf}$ it reaches the high frequency
plateau being determined by $\langle W\rangle$ (see
Sec.~\ref{section:model}).  In real materials this high frequency
plateau is usually masked by phonon excitations, but may be resolved
by substracting the vibrational contributions.
\cite{Cramer/etal:1995} The dispersive part of the spectrum for
$\tau^{-1}\ll\omega\ll\omega_{\rm hf}$ can be described by an
approximate power law at sufficiently low temperatures
($\sigma_{\varepsilon}/k_{\rm B} T\lesssim 4$, see
Fig.~\ref{figure:sqrr+sigma}b) with typical exponents $s$ in the range
0.5-0.8.  The imaginary part $\sigma''(\omega)$ behaves similar to
$\sigma'(\omega)$ in the dispersive regime $\tau^{-1} \ll \omega \ll
\omega_{\rm hf}$, i.e.\ $\sigma''(\omega) \sim \omega^s$, while in the
dc- and high frequency limits it rapidly approaches zero, since the
phase shift between current and electric field approaches rapidly $0$
and $\pi$, respectively.

\end{multicols}
\begin{multicols}{2}

\begin{figure}[h]
\hspace*{-0.3cm}\epsfig{file=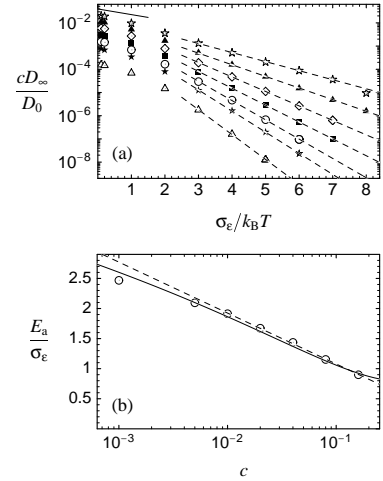,height=9.5cm}
\vspace*{-1.3cm}
\caption{(a)~Plot of $c D_{\infty}/D_0$ vs $\sigma_{\varepsilon}/k_{\rm B} T$
for concentrations $c = 0.001$ (open triangles) $0.005$ (full stars), $0.01$
(open circles), $0.02$ (full squares), $0.04$ (open diamonds), $0.08$ (full
triangles), and $0.16$ (open stars). The solid line indicates the
high-temperature activation energy \protect\cite{Note5} $E_0 =
\sigma_{\varepsilon}/\pi^{1/2}$. The dashed lines are the fits used to
determine the activation energy $E_{\rm a}$. (b)~Plot of the activation energy
$E_{\rm a}/\sigma_{\varepsilon}$ vs $c$. The open circles represent the values
obtained from the Monte Carlo simulation shown in (a), while the solid line
marks the result from the critical path analysis. The dashed lined is a fit
with respect to a logarithmic dependence of $E_{\rm a}$ on $c$, $E_{\rm a} =
A-B\log(c)$ with $A\protect\cong 0.23$ and $B\protect\cong 0.37$.}
\label{figure:activation}
\end{figure}

Figure~\ref{figure:activation}a displays Arrhenius plots of
$\sigma_{\rm dc}T\propto cD_{\infty}/D_0$ for various $c$. The linear
behavior of the curves at low $k_{\rm B} T\ll\sigma_{\varepsilon}$
(see the dashed lines in the figure) allows us to determine
\cite{Note5} the activation energy $E_{\rm a}$ whose dependence on $c$
is shown in Fig.~\ref{figure:activation}b (in a semi-logarithmic
plot).  This dependence is in good agreement with the behavior
predicted by the critical path analysis (see the solid line in
Fig.~\ref{figure:activation}b). As indicated by the dashed line in
Fig.~\ref{figure:activation}b, the data for $c\gtrsim 0.01$ can be
approximated by a logarithmic dependence $E_{\rm
  a}/\sigma_{\varepsilon} = A-B\log c$, in accordance with the
experimental findings. We also see that for low concentrations
$c\lesssim0.01$, the data start to fall off the straight dashed line.
This is expected to occur in view of the low-$c$ limit of
$\varepsilon_{\rm f}(c)$ discussed after eq.~(\ref{ea-eq}). The
behavior is not in contradiction with the experimental situation,
however, since in ionically conducting glasses the logarithmic
$c$-dependence of $E_{\rm a}$ is typically found at relatively high
ionic concentrations (corresponding to network modifier contents of
5-40 mol~\%, which are most interesting for applications).

\begin{figure}[t]
\vspace*{-4cm}
\hspace*{-1cm}\epsfig{file=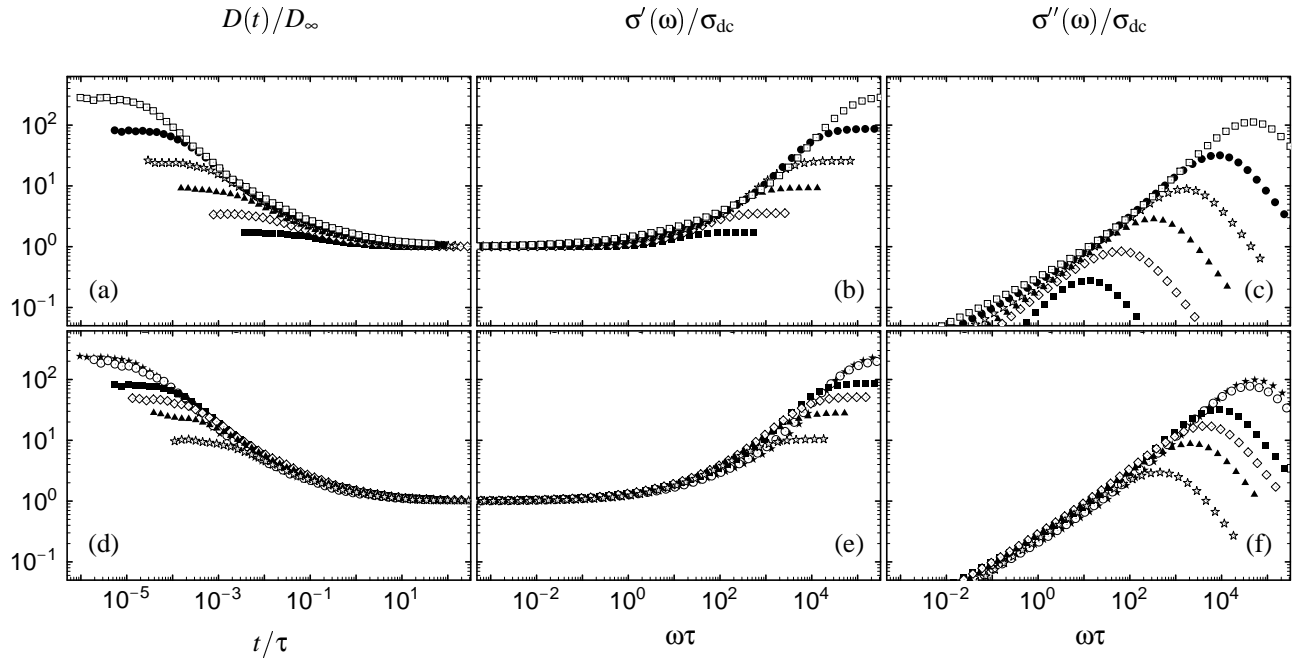,height=20cm,angle=-90}
\vspace*{-3cm}
\caption{Scaling plot of (a),(d)~$D(t)/D_{\infty}$ vs $t/\tau$,
(b),(e)~$\sigma'(\omega)/\sigma_{\rm dc}$ vs $\omega \tau$, and
(c),(f)~$\sigma''(\omega)/\sigma_{\rm dc}$ vs $\omega \tau$, for the data
presented in Fig.~\protect\ref{figure:sqrr+sigma}.}
\label{figure:scaling1}
\end{figure}

\noindent
\begin{minipage}{\hsize}\hspace*{0.2cm}
Let us now discuss as to what extent the data can be superimposed by a BNN
scaling ansatz as in eq.~(\ref{eq:bnn-scaling}). To this end the data in
Fig.~\ref{figure:sqrr+sigma} are plotted in a scaled form in
Fig.~\ref{figure:scaling1}. As can be seen from Figs.~\ref{figure:scaling1}a-c,
the data for fixed $c$ and various $T$ do not collapse onto one master curve.
The BNN scaling with respect to various concentrations at fixed (low)
temperature $T$ is significantly better (Figs.~\ref{figure:scaling1}d-f), but
it is important to note that this approximate scaling behavior ceases to be
present at higher
\end{minipage}
\vspace*{-1.8cm}
\vspace*{-0.5cm}
\end{multicols}
\begin{multicols}{2}
\noindent
$T$, $\sigma_{\varepsilon}/k_{\rm B}T\lesssim5$. We conclude
that the hard core lattice gas with Gaussian site energy disorder in general
exhibits no BNN scaling.

\begin{figure}[b]
\hspace*{-1cm}\epsfig{file=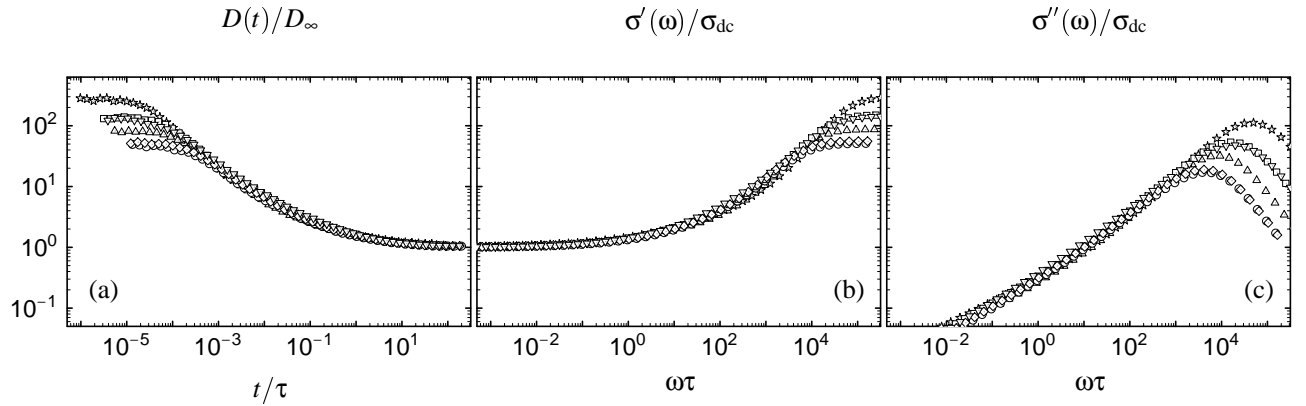,height=20cm,angle=-90}
\vspace*{-4.5cm}
\caption{Scaling plot of (a)~$D(t)/D_{\infty}$ vs $t/\tau$,
(b)~$\sigma'(\omega)/\sigma_{\rm dc}$ vs $\omega \tau$, and
(c)~$\sigma''(\omega)/\sigma_{\rm dc}$ vs $\omega \tau$, for $(c,
\sigma_{\varepsilon}/k_{\rm B} T) = (0.02,6)$ (open triangles), $(0.02,7)$
(open stars), $(0.04,6)$ (open circles), $(0.04,7)$ (open squares), $(0.08,7)$
(open diamonds), and $(0.08,8)$ (open upside down triangles).}
\label{figure:scaling2}
\end{figure}

However, the data collapse for the lowest temperatures
($\sigma_{\varepsilon}/k_{\rm B} T = 6$ and $7$) in
Fig.~\ref{figure:scaling1}a-c is fairly good. In order to see if this holds
true also for other $c$ values, let us first estimate what is a typical range
of ionic concentrations used in the experiments. For this purpose one might
consider sodium silicate glasses $x {\rm Na}{}_2 {\rm O} \cdot (1-x) {\rm Si}
{\rm O}{}_2$ with modifier contents $x$ between $10\%$ and $40\%$. By assuming
about $10$ sites per silicon to exist, we would have $c = 2 x/[10 (1-x)] \cong
0.01$ and $c \cong 0.13$ for $x = 0.1$ and $x = 0.4$, respectively. For these
$c$ values we always found
an approximate BNN scaling for $\sigma_{\varepsilon}/k_{\rm
  B}T\gtrsim6$.
A scaling plot combining various low $T$ and typical
$c$ values is shown in Fig.~\ref{figure:scaling2}. The quality of the
approximate data collapse is not as good as the one found in
experiment, but might become better at lower $T$. In fact, when
considering typical activation energies for ionic conduction in
glasses of $0.3$ to $1 \; {\rm eV}$, one should assume similar values
for $\sigma_{\varepsilon}$ (see Fig.~\ref{figure:activation}b), and
accordingly $\sigma_{\varepsilon}/k_{\rm B} T \le 10$ at room
temperature (this, however, is true only when neglecting the Coulomb
interaction between the mobile ions \cite{Maass/etal:1996}).
\end{multicols}
\vspace*{-5.5cm}
\begin{multicols}{2}

\section*{Acknowledgments}

We thank K.~Funke, M.~D.~Ingram, and B.~Roling for very helpful discussions.
M.P.\ and P.M.\ gratefully acknowledge financial support from the
Alexander-von-Humboldt foundation (Feodor-Lynen program) and the Deutsche
Forschungsgemeinschaft (Ma~1636/2-1), respectively.

\end{multicols}

\end{document}